# Matching Known Patients to Health Records in Washington State Data

Latanya Sweeney

Harvard University
latanya@fas.harvard.edu

*The State of Washington sells patient-level health data for $50. This publicly available dataset has virtually all hospitalizations occurring in the State in a given year, including patient demographics, diagnoses, procedures, attending physician, hospital, a summary of charges, and how the bill was paid. It does not contain patient names or addresses (only ZIPs). Newspaper stories printed in the State for the same year that contain the word "hospitalized" often include a patient's name and residential information and explain why the person was hospitalized, such as vehicle accident or assault. News information <u>uniquely and exactly</u> matched medical records in the State database for 35 of the 81 cases (or 43 percent) found in 2011, thereby putting names to patient records. A news reporter verified matches by contacting patients. Employers, financial organizations and others know the same kind of information as reported in news stories making it just as easy for them to identify the medical records of employees, debtors, and others.*

## INTRODUCTION

Imagine your personal health information flowing seamlessly between organizations and devices as needed. This liquidity in personal health information is to be the norm by 2015 according to the American Recovery and Reinvestment Act of 2009 (ARRA or the stimulus bill). ARRA provides financial compensation to healthcare providers and hospitals for using electronic health records [1, 2] so that patient measurements, diagnoses, procedures, medications, and demographics, along with physician notes and lab results will no longer be stored on paper but in digital format, enabling rapid and widespread sharing of patient data for many worthy purposes. To facilitate data sharing, ARRA supports data exchanges through State and regional repositories and networks and establishes comprehensive databases. Several ARRA initiatives go online next year. A broad range of concerns about these initiatives has been raised previously (e.g., [3]). A critical question is patient protection. *Are privacy safeguards sufficient to protect patients from harms*?

A concrete way to answer this question to test how well States currently protect patients when sharing health data widely beyond the care of the patient.

Years earlier, most States passed legislation requiring information about each hospital visit to be reported to the State, and most of those States in turn, share a copy of the information widely for many purposes. Actually, anyone can usually get a public version of the data that includes patient demographics, clinical diagnoses and procedures, a list of attending physicians, a breakdown of charges, and how the bill was paid for each hospitalization in the State. The information does not contain patient names but often includes residential postal codes (ZIPs). In comparison, data from ARRA initiatives will have these fields and many additional details, such as lab values and medical measurements, and will include office visits as well as hospitalizations.

Often medical information is benign –a broken arm gets a cast –but other times a hospitalization can include surprising results, such as drug or alcohol dependency appearing in an emergency hospitalization following a motor vehicle accident. So, care must be taken when sharing patient information.

Statewide databases have been around for years and shared widely. If there was a problem, one might expect to be able to point to a litany of harms, but a lack of enforcement and a lack of transparency confound findings. The Washington Post reported that the federal government received nearly 20,000 allegations of privacy violations, but imposed no fines and prosecuted only two criminal cases by 2006 [4]. As of this writing, I found no reported privacy violations from any State databases though it is unclear how and to whom one would report a violation. Most people are unaware of these statewide datasets, so even if harmed, it is unlikely a person would be able to link harms back to the shared data.

On the other hand, there is anecdotal evidence. In a 1996 survey of Fortune 500 companies, a third of the 84 respondents said they used medical records about employees to make hiring, firing and promotional decisions [5]. There have been allusions to a banker crossing medical information with debtor information at his bank, and if a match results, tweaking creditworthiness accordingly [6]. True or not, it is certainly possible, and the lack of transparency in data sharing makes detection virtually impossible even though the harm can be egregious. What is needed is a concrete example of how patients can be identified in this kind of data.

*If you know someone who went to the hospital and you know the approximate reason and/or the person's basic age, gender, and ZIP code, can you find his medical record in a State database?*

At first glance, identifying patients to medical records may seem academic or a question of curiosity. But if successful, employers might already be checking on employees' health, financial institutions adjusting credit worthiness based on medical information, data mining companies constructing personal medical dossiers from pharmacy records, and people snooping on friends, family and neighbors. Any of these parties could know when a person may have gone to the hospital and other relevant information needed to find that person's medical record.

## BACKGROUND

The Health Information Portability and Accountability Act (the first) in the United States is the federal regulation that dictates sharing of medical information beyond the immediate care of the patient, prescribing to whom and how physicians, hospitals, and insurers may share a patient's medical information broadly. State data collections are exempt from HIPAA. A State may share data mandated by State legislation in any form it deems appropriate. *How do State decisions compare to HIPAA*?

The Safe Harbor provision of the HIPAA Privacy Rule prescribes a way to share medical data publicly. Dates only report the year, and ZIP codes are the first 3 digits, if the population in those ZIP codes is greater than 20,000 and '00000' otherwise. No explicit identifiers, such as name, Social Security numbers, or addresses can appear.[1]

In comparison, many States share health information with less than a year specification on admissions and discharges and in some cases, providing the month and year of birth, not just the year [7]. Other states generalize values beyond the HIPAA standard, typically providing age ranges and/or ranges of ZIP codes. What guidelines should ARRA initiatives follow? The answer is not clear without specific examples of whether patients can be matched by name to the records.

---

[1] 45 CFR 164.514(b)(2) (2002).

Re-identifications of health data involve uniquely and specifically matching a named person to his medical record, and re-identifications have been done previously. In 1997, I showed how demographics appearing in medical data that did not have the names of patients could be linked to registries of people (e.g., voter lists) to restore name and contact information to the medical data [8]. My earliest example was identifying the medical information of William Weld, former governor of Massachusetts, using only his date of birth, gender, and 5-digit ZIP code appearing in medical data and a voter list [9]. I also used populations reported in the U.S. Census to predict that at most 87 percent of the U.S. population had unique combinations of date of birth, gender, and 5-digit ZIP [8].

Recently, others have challenged whether there really is any vulnerability to being re-identified by these fields, citing a lack of documented examples and of verified results (e.g. [10]). To address these concerns, new experiments are underway. Last month, a significant number of names and contact information were verified as being correctly matched to publicly available profiles in the Personal Genome Project using date of birth, gender and 5-digit ZIP [11]. Of course today, unlike the era of the Weld re-identification, State databases do not share full dates of birth, begging the question be investigated further. *Can patients be re-identified in today's State health data?*

## MATERIALS

Materials are: a collection of old news stories; an online public records service for locating basic demographics on Americans; and, a State database of hospitalizations in the same year and state as the news stories. More information about each resource appears below followed by descriptions of preliminary processing of diagnosis and hospital codes and admissions dates.

**News Stories**

Searching the LexisNexis newspaper archive [12] for news stories printed in 2011 in Washington State newspapers that contain the word "hospitalization" and that refer to a hospitalization of a person in 2011 yielded 66 distinct news stories from four news sources: Spokesman Review (28 stories), The Associated Press and Local Wire (17 stories), The Columbian (19 stories), and Mukilteo Beacon (2 stories). Table 1 has a summary.

Figure 1 provides an example news story about a motorcycle crash in which 61-year-old Raymond Boylston, from Soap Lake, Washington, is sent to Lincoln Hospital.

The news stories referenced a total of 111 people. Some stories described incidents involving multiple people. Not all stories contained a name of a person; only 86 names appeared in the news stories. One story did not have the names of the people, but did have an explicit street address in which 4 people who lived there were hospitalized following a house fire. So, the total number of subjects is 90, which is the 86 named people and the 4 people residing at the known street address.

Most news stories that listed names were of motor vehicle crashes (51 stories) and assaults (12 stories). Some stories reported medical hospitalizations (13 stories), primarily of popular people (e.g., a professional soccer player, a judge, and a Congressman). The remaining 14 stories reported shootings, suicide attempts, house fires, and other events. Table 2 lists types of stories for the 90 subjects.

News stories tend to report the person's *name*, *age*, *residential information*, *type of incident*, *incident date*, and *hospital*.

|  | Number of News Stories |
|---|---|
| Spokesman Review (Spokane, WA) | 28 |
| The Associated Press & Local Wire | 17 |
| Mukilteo Beacon | 2 |
| The Columbian (Vancouver, WA) | 19 |
| Total | 66 |

**Table 1. Distribution of news stories by news source for a total number of 66 stories.**

> MAN, 61, THROWN FROM MOTORCYCLE
> A 61-year-old Soap Lake man was hospitalized Saturday afternoon after he was thrown from his motorcycle.  Raymond Boylston was riding his 2003 Harley-Davidson north on Highway 25, when he failed to negotiate a curve to the left. His motorcycle became airborne before landing in a wooded area. Boylston was thrown from the bike; he was wearing a helmet during the 12:24 p.m. incident.  He was taken to Lincoln Hospital.
> [Spokesman Review 10/23/2011]

**Figure 1. Sample extract of a news story that contains *name*, *age*, *residential information*, *hospital*, *incident date*, and *type of incident*.**

**NEWS STORIES**

|  | Number of Subjects | Percent |
|---|---|---|
| Motor Vehicle | 51 | 57% |
| Assault | 12 | 13% |
| Medical | 13 | 14% |
| Other | 14 | 16% |
| Totals | 90 |  |

**Table 2. Distribution of news stories by type of incident for 90 subjects.**

Harvesting these values, as available, from the news stories and adding the news source and publication date comprise the "NewsData" dataset used in this study, which starts with 90 records, one for each subject. Figure 2 reports the distribution of fields in NewsData. Gender is present in all the records. Seventeen records have all the fields and 31 records have six of the fields. Only one record has just name, gender and address with no hospital or medical content.

**Online Public Records**

Numerous online services offer search facilities for government-collected information (or public records) in the United States about a person. When a person's name and/or other demographics are entered, these services may return the person's date of birth, history of residential addresses, phone numbers, criminal history, and professional and business licenses, though specifics vary among states and services. Prices and results vary too. Some are free online, but most services offer a per lookup fee (e.g., $3 to $10 per lookup). Some services offer monthly or yearly subscription plans, which can significantly lower per-lookup costs (e.g., $0.75 per lookup for up to 300 searches, or unlimited searches for $40/year). In this study, references to these kinds of search results are called "PublicRecords".

**Hospital Data**

The "Comprehensive Hospital Abstract Reporting System: Hospital Inpatient Dataset: Clinical Data" [13] lists hospitalizations in Washington State for the Year 2011 and costs $50. This data contains a record for each hospitalization in the State and the total number of hospitalizations (or records) is 648,384. Each hospitalization is described using 88 fields of data; these include: patient's 5-digit ZIP code, age in years and months, race, ethnicity, and gender; hospital; month of discharge; number of days in the hospital; admission type, source and weekend indicator; discharge status; how the bill was paid; diagnosis codes; procedure codes; and list of attending physicians. This dataset is termed "HospitalData" in this study.

For computer storage efficiency, most fields contain codes rather than English descriptions, so Washington State provides a dictionary that defines each code.

| Number of Fields | | Name or Street | Gender | Type | Age | General Address | Hospital | Details | Number of Subjects | Totals |
|---|---|---|---|---|---|---|---|---|---|---|
| 3 | | ■ | ■ | | | ■ | | | 1 | 1 |
| 4 | a | ■ | ■ | ■ | | | | ■ | 5 | 14 |
| | b | ■ | ■ | ■ | ■ | | | | 7 | |
| | c | ■ | ■ | ■ | | ■ | | | 1 | |
| | d | ■ | ■ | | ■ | ■ | | | 1 | |
| 5 | a | ■ | ■ | ■ | ■ | | | ■ | 6 | 27 |
| | b | ■ | ■ | ■ | | ■ | | ■ | 7 | |
| | c | ■ | ■ | ■ | ■ | ■ | | | 4 | |
| | d | ■ | ■ | ■ | | | ■ | ■ | 6 | |
| | e | ■ | ■ | ■ | ■ | | | ■ | 3 | |
| | f | ■ | ■ | ■ | | ■ | | ■ | 1 | |
| 6 | a | ■ | ■ | ■ | ■ | | ■ | ■ | 4 | 31 |
| | b | ■ | ■ | ■ | ■ | ■ | | ■ | 9 | |
| | c | ■ | ■ | ■ | ■ | | ■ | ■ | 17 | |
| | d | ■ | ■ | ■ | | ■ | ■ | ■ | 1 | |
| 7 | | ■ | ■ | ■ | ■ | ■ | ■ | ■ | 17 | 17 |
| | | | | | | | | Totals | 90 | 90 |

Figure 2. Distribution of values for fields harvested from news stories. Name is present in 86 cases with 4 others having an explicit street address, for a total of 90 subjects. "General Address" refers to generalized residential information, such as town, city, county or region of the state. "Type" and "Details" refer to the kind of incident and any medical details.

Figure 3 provides an example of part of a record in HospitalData, showing the code and its definition where appropriate. It describes an emergency admission (Admit Type field) of a 60-year-old, 725-months-old (Age fields) white, non-Hispanic (Race/Ethnicity) male (Gender). The total charge of $71,708 and was paid by a combination of three payers. The emergency is from a motorcycle accident (Emergency Codes) and the diagnoses include a fracture of his pelvis (Diagnosis Codes).

**Diagnosis Codes to NewsData**

In a preliminary step, adding diagnosis fields to NewsData makes it more compatible for matching to HospitalData.
Using the type of incident described in a news story, we can list diagnosis codes that would likely appear among the diagnosis codes in HospitalData. The International Classification of Diseases 9th edition (or ICD9 codes) defines more than 15,000 diagnosis codes [14], grouped into three categories: all numeric codes describe medical diseases (e.g., diseases of the digestive system or complications of pregnancy); codes beginning with an 'E' describe external causes of injury or poisoning (e.g., motor vehicle accidents or assaults), and codes beginning with a "V" describe factors that may influence health status (e.g., communicable diseases, drug dependency, or tobacco use).

An ICD9 diagnosis code is an alphanumeric string where the leftmost characters represent a family of values, made more specific by adding more characters to the right. Figure 4 shows an example using emergency codes that begin with E81, which are some motor vehicle accident codes. A code of E816 describes an accident involving an out of control motor vehicle that had no collision. Adding another digit provides more detail about how the patient was involved or injured. For example, E8162 describes the case where the vehicle was a motorcycle and the motorcyclist was injured.

More than half of the news stories involved motor vehicle accidents and a substantial number of stories described assaults (see Table 2). The ICD9 codes for motor vehicle accidents begin with the codes E81 and E82. Assaults begin with E96. Table 3 reports the numbers of records containing these codes in HospitalData: 5232 motor vehicle accidents and 1612 assaults.

| | |
|---|---|
| Hospital | 162: Sacred Heart Medical Center in Providence |
| Admit Type | 1: Emergency |
| Type of Stay | 1: Inpatient |
| Length of Stay | 6 days |
| Discharge Date | Oct-2011 |
| Discharge Status | 6: Dsch/Trfn to home under the care of an health service organization |
| Charges | $71708.47 |
| Payers | 1: Medicare |
| | 6: Commercial insurance |
| | 625: Other government sponsored patients |
| Emergency Codes | E8162: motor vehicle traffic accident due to loss of control; loss control mv-mocycl |
| Diagnosis Codes | 80843: closed fracture of other specified part of pelvis |
| | 51851: pulmonary insufficiency following trauma & surgery |
| | 86500: injury to spleen without mention of open wound into cavity |
| | 80705: closed fracture of rib(s); fracture five ribs-close |
| | 5849: acute renal failure; unspecified |
| | 8052: closed fracture of dorsal [thoracic] vertebra without mention of spinal cord injury |
| | 2761: hyposmolality &/or hyponatremia |
| | 78057: tachycardia |
| | 2851: acute posthemorrhagic anemia |
| Age in Years | 60 |
| Age in Months | 725 |
| Gender | Male |
| ZIP | 98851 |
| State Reside | WA |
| Race/Ethnicity | White, Non-Hispanic |
| Procedure Codes | 5781: Suture bladder laceration |
| | 7939: 7919: Open/Closed reduction of fracture of other specified bone |
| Physicians | … |
| … | … |

**Figure 3. Sample extract of fields of information from a hospitalization record in HospitalData.**

| | |
|---|---|
| 001 | Cholera |
| | 0010  …due to vibrio cholerae |
| … | … |
| E810 | Motor vehicle traffic accident involving train collision |
| | E8100  …injuring driver of vehicle not motorcycle |
| | E8101  …injuring passenger in vehicle not motorcycle |
| | E8102  …injuring motorcyclist |
| | E8103  …injuring passenger on motorcycle |
| … | … |
| E816 | Motor vehicle accident, loss control highway not collide |
| | E8160  …injuring driver of vehicle not motorcycle |
| | E8161  …injuring passenger in vehicle not motorcycle |
| | E8162  …injuring motorcyclist |
| | E8163  …injuring passenger on motorcycle |
| | E8164  …injuring occupant of streetcar |
| | E8165  … injuring rider of animal; animal-drawn vehicle |
| | E8166  …injuring pedal cyclist |
| | E8167  …injuring pedestrian |
| | E8168  …injuring other specified person |
| | E8169  …injuring unspecified person |
| E817 | Noncollision motor vehicle traffic accident while boarding |
| … | … |
| E999 | Late effect of injury due to war operations and terrorism |
| | E9990  Late effect of injury due to war operations |
| | E9991  Late effect of injury due to terrorism |
| V01 | Exposure to communicable diseases |
| … | … |
| V91 | Multiple Gestation Placenta Status |

**Figure 4. Excerpt of ICD9 Diagnosis coding tree. As more digits get appended to the right, the details get more specific.**

**HOSPITAL DATA**

| | Number of Records | Percent |
|---|---|---|
| Motor Vehicle | 5232 | 0.8% |
| Assault | 1612 | 0.2% |
| All others | 641540 | 98.9% |
| Totals | 648384 | |

**Table 3. Number of health records having a diagnosis starting with E81 or E82 for motor vehicle accidents and E96 for assaults from a total of 648,384 hospitalizations in HospitalData.**

To facilitate simple matching of NewsData to HospitalData, I added diagnosis fields to NewsData and populated the fields with general versions of ICD9 codes (the leftmost digits) whose descriptions matched the incident details harvested from the news stories (see Details in Figure 2). In the 51 cases involving motor vehicle accidents, I merely recorded "E81" and "E82". In the 12 assault cases, I recorded "E96". In the remaining 27 cases, I used automated search for ICD9 codes whose descriptions matched details harvested from news stories. I recorded the most general version of the ICD9 code that matched the description – i.e., the three leftmost characters only.

For example, one news story reported Congressman Hastings was hospitalized with "diverticulitis". A search of this term yielded ICD9 code "56211", so the code "562" was added to his record in NewsData.

As shown in Figure 2, all but two of the 90 records in NewsData had content in the incident Type or Details field. I matched news descriptions to ICD9 codes in 72 of the 90 cases (or 80 percent of the stories) in NewsData. The specific diagnosis codes were: 437, 444, 508, 518, 562, 569, 800, 801, 802, 803, 804, 805, 808, 818, 824, 827, 829, 861, 864, 873, 884, 900, 910, 920, 923, 942, 943, 944, 945, 946, 947, 959, V58, E81, E82, E88, E89, E92, E95, E96, E97, and E98.

**Hospital Codes to NewsData**

Many of the news stories included the name of the hospital (50 of 90 or 56 percent, as listed in Figure 2). One news story merely stated a "Tri-Cities hospital" which is one in a group of about a dozen possible hospitals. As described earlier, HospitalData uses codes instead of English text in many fields, and one such field is *hospital*. An example appears in Figure 3. The code for Sacred Heart Medical Center in Providence is 162. Lincoln Hospital has a code of 137.

A dictionary of hospitals was included with HospitalData that lists 183 hospitals in Washington State along with their assigned codes. Some hospitals appear multiple times, with a letter added to the code to distinguish different units (e.g., rehabilitation or acute care).

The name of each hospital appearing in a news story was automatically looked up in the dictionary of Washington State hospitals and a new field added and populated that had the code for the Hospital.

Nine of the hospitalization reports in the news stories were in other states, specifically Oregon and Idaho, and as a result, those hospitalizations would not be in HospitalData. Therefore, these records were removed from NewsData, lowering the number of actual subjects to 81. *The total number of subjects in the remainder of this writing will be 81 unless otherwise stated.*

Other than those news stories referencing out-of-state hospitals and the one news story that referenced a Tri-City hospital, all other hospitals present in the news story were uniquely matched and appended to NewsData.

**Admission Dates to HospitalData**

HospitalData includes the month and year of the discharge and the length of stay in days but has no field for the admission date. On the other hand, a new story reports when an incident occurred. The date of the incident in the new story corresponds to the hospital admission date. So, I use the month of the discharge and the number of days in the hospital to compute a one-month range for the admission.

The earliest possible day of the admission would be the first day of the month of the discharge less the number of days in the hospital ("*admit begin*"). The latest possible day of the admission would be the last day of the month of the discharge less the number of days in the hospital ("*admit end*"). The date of the incident reported in the news story must be on or after *admit begin* but before or on *admit end* for the news story to match that record of HospitalData.

For example, the news story in Figure 1 has an incident date of October 18, 2011. The medical record in Figure 3 reports a discharge month of October 2011 and a 6-day stay. The earliest the admission could have occurred was September 25, 2011 (admit begin) and the latest was October 25, 2011 (admit end). So, the news story incident date matches the possible admission date for the medical record.

I have now extended NewsData to include diagnosis codes and hospital codes based on information appearing in the news stories, and reduced the number of records in NewsData from 90 to 81 by discarding news reports about out of state hospitalizations. I have also extended HospitalData to include two fields, *admitbegin* and *admitend*, that describe a one month window in which the admission must have occurred. Armed with these enhancements, we are ready to match NewsData with HospitalData.

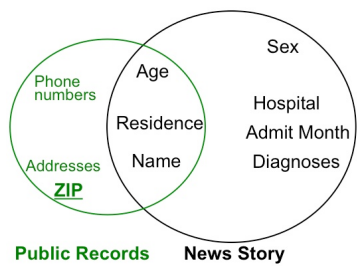

**Figure 5. Acquire 5-digit ZIP codes from public records using {name, residence information, age} from the news story. Age in years is from news and date of birth from public records.**

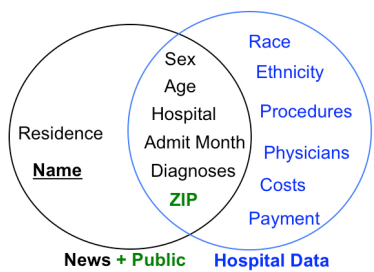

**Figure 6. The automated approach matches news information and ZIP (from public records) to hospital data on a combination of {gender, age, hospital, admit month, diagnoses related to incidence, ZIP}, thereby putting a name to a medical record. Age is in years and months and the month of birth comes from public records.**

**APPROACH**

As described earlier, a news story containing the word "hospitalization" often includes some combination of *name*, *age*, *residence*, *gender*, *hospital*, and *incident* specifics sufficient to infer admission month and some of the diagnoses. An initial step is to look up the person's *name*, *age* and *residence* information in PublicRecords to learn the person's *date of birth* and any 5-digit *ZIP* codes associated with the person. (This step can be automated with the purchase of a public records database.) Figure 5 provides a depiction of this initial step. Afterwards, two different approaches were investigated, one involved automated matching of fields and the other uses human exploration.

**Automated Approach**

A direct computer comparison is made between the newly learned *ZIP* and *age in months* (derived from birthdate and incident date) and other information from the news article –i.e., *gender*, *age*, *hospital*, *admission month*, and some likely *diagnoses*– to the fields of information in each record in HospitalData. Any blank fields in NewsData are not used for comparison. When the overall comparison uniquely matches on each field of information provided, and the match of a news story is to one and only one record in HealthData, the result associates the *name* and *residence* information from the news story to the person's health data, even though the health data did not previously include the name of the patient. If the comparison relates a news story to more than one record in HealthData, the comparison does not yield a match in this study. Only exact and unique matches are accepted. If no match is found, the approach repeats suppressing one or two values in the NewsData to see if one and only matching record appears in HealthData.

For example, the news story in Figure 1 contains {Raymond Boylston, 60 year old, male, from Soap Lake, admitted to Lincoln Hospital, in October 2011, for a motorcycle-related accident}. If a search of HospitalData yields a unique match on these fields, then I consider the result a match of the news story to the hospital record. If no match is found, one field in the news story is dropped and tested, and the test repeats dropping each field in turn. If no records uniquely match when any one field is dropped, then two fields are dropped and each possibility tested. Figure 2 shows a uniquely matching health record when age and hospital are dropped.

**Human Approach**

A temporary employee, who was resourceful and knowledgeable about how to use the Internet and computers but who had no degree or training in computer science, mathematics, statistics or medicine, was hired through an employment agency as a "human investigator". I did not know her previously. After reviewing some of the matched and unmatched cases resulting from the automated approach, I asked the temp to see if she could find matching records for some of the new stories missed by the automated approach –i.e., cases where using all the fields in the news story led to no match. She could only use a set of candidate medical records, the news story, and any information she found publicly and freely available on the Internet.

**Scoring**

A news reporter agreed to score results by interviewing those matched to confirm the contents of the health record with the person. (No medical information or personal identity would be publicly revealed unless the person explicitly agreed to share his identity or medical information publicly.)

**RESULTS**

Directly matching the fields of the records in NewsData to those in HealthData yielded unique and exact matches on 35 of the 81 records in NewsData (or 43 percent). Ten of the records in NewsData matched two records in HospitalData ambiguously, eleven matched three or more records in HealthData, and 25 matched none. The matches were systematically reviewed for consistency with details in the news story and other online information and no inconsistencies were found.

Of the 35 exact and unique matches, 30 matches used all the values supplied in the news story, 5 matches resulted when one value from the news story was dropped (ZIP, age, or hospital) and 1 match resulted when two values were dropped (age and hospital). In one case, the name that appeared in the news story was not found, but a public records search without the last name matched one person exactly and that name with her associated ZIP and age was used.

The scientific accuracy of these results rests in the accuracy of the newspaper, health and public records information. Hospitals in the State must provide the information to the State by law, and the fields are drawn directly from billing records. HealthData reportedly contains all hospitalizations in the State [13]. Most of the news articles are from police reports; one was from a press release of the hospitalization of a Congressman and a few seemed to be regular news stories. There is no guarantee that these data are error free, but any errors made in matching must result from errors in the data sources and not in the matching.

The automated approach uses an exact match, not a probabilistic one. Probabilistic matching would likely increase the number of matches, but my interest is to get as accurate a match as the data allows, not as many matches as likely.

The news reporter was given the 35 matched results, from which he chose 15 to pursue by phone. He contacted 8, and confirmed all 8 were correctly matched (100% correct). He attempted to contact 6 others, but was not able to make phone contact with them during the 7-day evaluation period. Details from his interviews with those who agreed to public disclosure appear in his associated news story [15].

The human investigator was given two cases of high-profile people, a soccer player and a Congressman, one case for which the news story was unusual (a sky-diving accident), and two other cases. No match was found using automated comparison for these cases without dropping one or two values. The human investigator had two workdays in which to investigate these five cases.

She compiled portfolios on each case, documenting not only which record in the HealthData was correct, but also why the exact automated comparison failed to match. She was successful in all 5 cases.

In the two cases of high profile people, the ZIP code was not of a personal residence but the soccer franchise in the case of the soccer player, and his campaign headquarters in case of the Congressman. When these ZIP codes were used, the modified news information uniquely and exactly matched the same health records the human investigator found.

The news story of the sky diving accident found in LexisNexis had very few details. When she looked online at other news stories, she found his age and the fact that he lived in another state. Because he had an out of state ZIP code, she was able to quickly identify his record. When his personal information was appended to the news record, the automated comparison uniquely matched the same hospital record.

Similarly, her success in the other two cases resulted from augmenting the original news story with additional or correcting information she found online.

Most of the health records seemed to just report specifics one would expect to find related to the reported incident. But about one-third of the health records that were automatically matched to news stories (10 of 35 records) included references to venereal diseases, drug dependency, alcohol use, tobacco use, and other diagnoses or payment issues that may be sensitive, even though most of these records were for motor vehicle accidents.

## DISCUSSION

This experiment demonstrates how medical information for a targeted individual can be obtained using automated or human means, and neither takes sophisticated expertise. As one would expect, automated matching gives more results quickly than the human approach, but the human investigator located and used other sources specific to the record.

There are many more newspaper sources available in the State. This study only used those available through one service, but other news services offer other sources, and most newspapers have their own online websites. Overall, this means the number of possible cases drawn from news could be substantially larger.

Even though this study used newspapers as source information about a patient's identity, an employer could use hospital leave information to check on employees, a financial institution could use credit account information to assessing credit worthiness of clients who report medical illness as a reason for delayed payments, and a person snooping on friends or family could merely inquire about when, where and why the person was hospitalized. Even a data mining company could use prescription information to locate health records using the name, date of birth, address about the person, and the medicine and dosage to infer diagnosis codes.

This experiment is important because it demonstrates how health data, which is currently shared publicly and widely without the knowledge of the patient, could be putting patients at risk. This is a timely consideration as America gets ready to unleash health information exchanges, which will share much more health information on each patient and the data sharing will be far less transparent.

**What Can Be Done**

The goal is not to stop data sharing. On the contrary, sharing data beyond the patient encounter offers many worthy benefits to society. These data may be particularly useful because they contain a complete set of hospital discharges within the state, thereby allowing comparisons across regions and states such as rating hospital and physician performances and assessing variations and trends in care, access, charges and outcomes. Research studies that have used these datasets include: examinations of utilization differences based on proximity [16], patient safety [17, 18], and procedures [19]; and, a comparison of motorcycle accident results in states with and without helmet laws [20]. The very completeness that helps these studies makes it impossible to rely on patients to consent to sharing because the resulting data would not be as complete.

The goal is not to stop data sharing but to be smarter about how data sharing is done. This is a particularly important as the top buyers of statewide databases are not researchers, but private companies, especially those constructing data profiles on individuals [21].

Washington State could choose to share its data in a form that adheres to the standards set by the HIPAA Safe Harbor – reporting dates in years and geography in 3-digit ZIP codes (and blank if the ZIP has a small population). In patient demographic fields, Washington's data has 5-digit ZIP codes and age, appearing more protective than HIPAA, because an age is a 2-year range. However, Washington also shares the month and year of discharge and the number of days hospitalized, allowing inference of a month range for the admission date. Also, Washington's data has a field that gives the age of the patient in months, which when reversed, gives year and a 2-month window for the birth month. These extra inferences, 5-digit ZIP code, admission month and bi-monthly birth year, are more specific than just year, and were useful in matching.

To be equivalent to HIPAA, Washington State could drop the discharge month and reporting no more than the year of birth of the patient. Of course, these redactions may cause the resulting data to be less useful for some purposes, so for those uses, Washington State may have additional requirements for someone to meet in order to acquire the more sensitive data.

Better options promise to come from technology. The same technology that has brought us today's data rich networked society is the same technology that can provide the best privacy protection. To get there, however, we have to align policy and technology incentives, and that too is where this experiment fits in.

Policy should use best practices, which improve over time as privacy technology and the science of data privacy advances. Society can benefit from cycles of published re-identifications, because doing so will rapidly lead to improved techno-policy protections. It is an evolutionary cycle. First, a re-identification vulnerability becomes known, which leads to improved practices and technical solutions, which in turn, leads to more re-identifications, and so on, until eventually we achieve robust technical solutions.

The cyclic approach of expose-then-improve is how strong encryption developed. Today, we use strong encryption for all sorts of tasks, such as online banking and purchasing. But the earliest forms of encryption were just ad hoc decisions, similar to the kind of ad hoc decisions made about data sharing today. Then, someone would publish a way to break the leading scheme of the time, spawning others to develop better methods, which in turn, would be broken, until eventually, we got the strong encryption society enjoys today. That's the same kind of cycle we need for data privacy so that society can enjoy widespread data sharing with privacy protections and not be forced to falsely choose between privacy or the benefits of data sharing.

## Acknowledgments

The author gratefully thanks Lauren Mimran Ugbabe for her effort as a human investigator, Sean Hooley and Ryan Joyce for carefully and methodically reviewing materials, Jordan Robertson for validating results, Raymond Boylston for allowing his information to be shared publicly, and Dr. Deborah Peel at Patient Privacy Rights for inspiring this study. A copy of data used in this study is available at foreverdata.org and the Dataverse Network.